\documentclass[pre,preprint]{revtex4-1}

\usepackage{amsfonts}
\usepackage{amsmath}
\usepackage{amssymb}
\usepackage{graphicx}
\usepackage{mathrsfs} 
\usepackage{color} 

\providecommand\bnabla{\boldsymbol{\nabla}}

\providecommand\bcdot{\boldsymbol{\cdot}}

\providecommand\br{\mathbf{r}}

\providecommand\bx{\mathbf{x}}

\providecommand\bd{\mathbf{d}}
\providecommand\bzero{\mathbf{0}}

\providecommand\be{\mathbf{\hat{e}}}
\providecommand\bbb{\mathbf{\hat{b}}}
\providecommand\bn{\mathbf{\hat{n}}}
\providecommand\bk{\mathbf{\hat{k}}}

\newcommand{\pd}[2]{\frac{\partial #1}{\partial #2}}

\newcommand{\bb}[1]{\boldsymbol{#1}}
\newcommand{\ub}[1]{^{({#1})}}

\newcommand{\TheTitle}{Asymptotic modelling of phononic box crystals}

\begin{document} 

\title{{\TheTitle}}
\author{A. L. Vanel, R. V. Craster \& O. Schnitzer}
\affiliation{Department of Mathematics, Imperial College London, London SW7 2AZ, UK}

\begin{abstract}
We introduce phononic box crystals, namely arrays of adjoined perforated boxes, as a three-dimensional prototype for an unusual class of subwavelength metamaterials based on directly coupling resonating elements. In this case, when the holes coupling the boxes are small, we create networks of Helmholtz resonators with nearest-neighbour interactions. We use matched asymptotic expansions, in the small hole limit, to derive simple, yet asymptotically accurate, discrete wave equations governing the pressure field. These network equations readily furnish analytical dispersion relations for box arrays, slabs and crystals, that agree favourably with finite-element simulations of the physical problem. Our results reveal that the entire acoustic branch is uniformly squeezed into a subwavelength regime; consequently, phononic box crystals exhibit nonlinear-dispersion effects (such as dynamic anisotropy) in a relatively wide band, as well as a high effective refractive index in the long-wavelength limit. We also study the sound field produced by sources placed within one of the boxes by comparing and contrasting monopole- with dipole-type forcing; for the former the pressure field is asymptotically enhanced whilst for the latter there is no asymptotic enhancement and the translation from the microscale to the discrete description entails evaluating singular limits, using a regularized and efficient scheme, of the Neumann's Green's function for a cube. We conclude with an example of using our asymptotic framework to calculate localized modes trapped within a defected box array. 
\end{abstract}
\maketitle

\section{Introduction}
\label{sec:intro}
The study of microstructured media in acoustic settings is undergoing a
considerable revival, with ideas originating from electromagnetism, photonic crystals and metamaterials 
influencing structured acoustic devices.  
Fundamentally, these advancements rely on incorporating resonances to effectively control wave propagation, localisation and attenuation. Building on a rich history of using acoustic resonators to influence sound propagation (e.g., in the context of filters \cite{Stewart:31}, bubbly liquids \cite{Carstensen:47} and sound transmission \cite{fahy87a}), the field of metamaterials has inspired numerous novel applications \cite{craster12a,deymier13a}, including ultrathin metasurface acoustic absorbers
for sound insulation \cite{cai14a,jimenez16b},
 media with negative effective properties \cite{fang06a,lee09a,liu00a}, 
  soda-can metamaterials \cite{lemoult17a} demonstrating sub-wavelength
lensing, and fishnet acoustic metamaterials for broadband tuneable
sound absorption \cite{christensen10a,murray14a}.

A common feature of metamaterial designs, that has been particularly influential, is that of using locally resonant elements whose dimensions are small compared to the operating wavelength. The capability to move to such subwavelength regimes distinguishes acoustic metamaterials from 
phononic crystals \cite{laude15a}, that rely on Bragg
 scattering and interference to achieve their performance. The most common subwavelength resonator used to realise phononic metamaterials is the Helmholtz resonator, namely a hollow vessel with a small opening  \cite{rayleigh1870}.
The Helmholtz resonator acts like a mass-spring system: the slug of fluid moving in the small opening acts as the mass, while the compressible fluid within the vessel generates a restoring force providing the spring action. 
 In the limit where the hole size is small compared to the dimensions of the vessel, classical scaling arguments show that 
 the resonant frequency
 lies in the subwavelength regime, scaling as ${\omega} = c\sqrt{S/(LV)}$,
 where $c$ is the speed of sound, $S$ is the neck
 area, $L$ is a characteristic neck length and $V$ is the volume of the vessel. 

Phononic metamaterials are usually
formed by arrays of physically separated subwavelength resonators,
any interaction between the resonators being mediated by an acoustically transparent background and often regarded as a nuisance. 
In this form, the metamaterial inherits its properties from its unit-cell element, typically in narrow frequency bands about resonant frequencies \cite{fang06a,griffin20}. 

Rather than considering such arrays of isolated resonators, one can envisage arrays of subwavelength resonators that are directly coupled. In a recent Letter \cite{vanel17a}, we studied wave phenomena in such media in two dimensions, specifically doubly periodic arrays of closely spaced rigid cylinders. We showed that, in the limit where the cylinders nearly touch, the voids of the structure, and the narrow necks connecting them, effectively form a network of interconnected Helmholtz resonators. Building on the classical lumped model of an isolated Helmholtz resonator, we developed discrete wave equations connecting the void pressures. Notably, the rapid pressure variations in the necks were resolved systematically, yielding an entirely analytical framework where the lumped parameters appear in closed form. The model reveals that the acoustic branch is uniformly squeezed into a subwavelength regime lying below a wide band gap. As a consequence, the crystal exhibits nonlinear dispersion in a relatively wide subwavelength regime, excluding the long-wavelength limit (i.e., well below cut-off) where it is characterized by a high effective index. Our asymptotic approach was shown to be in good agreement with finite-element simulations, and extremely versatile, allowing us to model and easily tune arbitrary lattice and inclusion geometries.
\begin{figure}[b!]
\begin{center}
\includegraphics[scale=0.6]{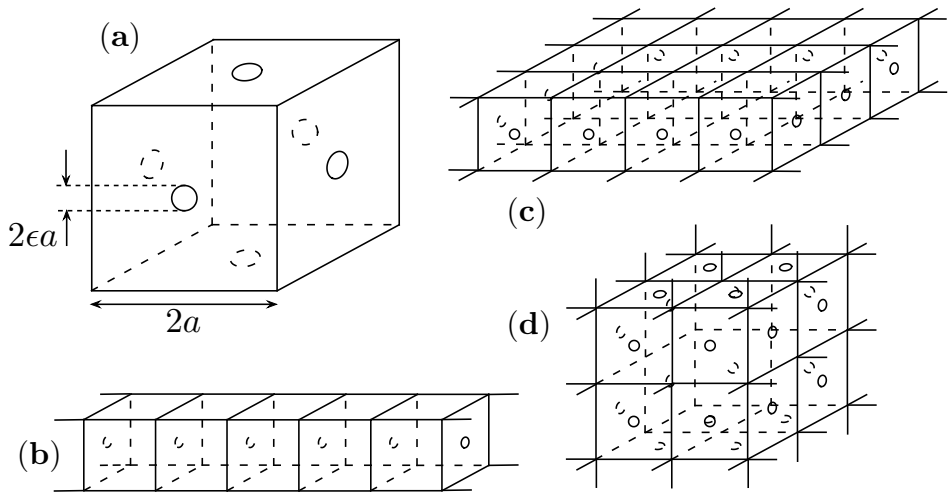}
\caption{A schematic showing (a) the unit box that forms the essential
  building block for the phononic box crystals in (b) as a waveguide,
  (c) as a slab array and (d) as a three-dimensional array.}
\label{fig:schematic}
\end{center}
\end{figure}

In this paper we consider what may be
the simplest three-dimensional example of a phononic metamaterial formed of interconnected Helmholtz resonators \footnote{A naive extension from two-dimensional arrays of closely spaced cylinders to 
three dimensions is to consider a cubic array of closely spaced spheres. In contrast to cylinder arrays, however,  sphere arrays exhibit linear dispersion throughout the subwavelength regime. Indeed, the fluid can flow between cells even when the spheres touch,  
hence the voids do not act as Helmholtz resonators.}: 
phononic box crystals,
namely arrays of adjoined perforated boxes, say
  cubic boxes forming
either triply periodic crystals, doubly periodic slabs or periodic waveguide arrays (see
Fig.~\ref{fig:schematic}).
Each constituent box 
has side length $2a$ and every face can be  
perforated at its center by a circular hole of radius $\epsilon a$; the walls are assumed infinitesimally thin in the mathematical analysis. Evidently, each perforated box is a Helmholtz resonator with effective dimensions 
$L=O(\epsilon a)$, $S=O(\epsilon^2 a^2)$ and $V=O(a^3)$, 
 suggesting a cut-off frequency $\omega a/c=O(\epsilon^{1/2})$; phononic box crystals can therefore be tuned, 
using the hole radii, to be
subwavelength.  In contrast, with this $O(\epsilon^{1/2})$ scaling, the cut off frequency for arrays of closely spaced cylinders 
\cite{vanel17a} scales 
as $O(\epsilon^{1/4})$, wherein $\epsilon$ is the width of the parabolic gaps relative to the pitch. 

Our aim here is to asymptotically study phononic box crystals in the small hole limit $\epsilon\ll1$, with the frequency assumed to be in the resonant regime $\omega a/c=O(\epsilon^{1/2})$. In doing so, we shall advance the
theory presented in \cite{vanel17a} in two directions. First, by going to three dimensions, where asymptotic modelling becomes essential owing to the immense computational challenge of accurately resolving large structures involving many hundreds or thousands of unit cells, each featuring several small holes. In particular, tractable network models as developed in \cite{vanel17a} enable rapid insight. We accordingly wish to show that  analogous models hold also for phononic box crystals and to derive these systematically using matched asymptotic expansions \cite{hinch91a}. Second, we shall go beyond extracting dispersion relations, and the Bloch eigenvalue problem, to consider forced problems where the sound field is induced by sources within the crystal. These sources clearly break the periodicity of the problem, which accordingly can no longer be reduced to a single unit cell using Bloch--Floquet theory. As we shall see, the asymptotic translation of a physical source on the microscale to equivalent forcing terms in the network approximation is subtle and depends crucially on the character of the applied forcing. 

The rest of the paper is structured as follows: We begin in \S\ref{sec:formulation} by formulating the homogeneous (Bloch-eigenvalue)  and forced problems. Then in \S\ref{sec:bloch} we analyze the homogeneous problem using matched asymptotic expansions, deriving network equations and dispersion relations describing the entire acoustic branch, which we compare with finite-element simulations. We next analyze the forced problem in \S\ref{sec:forced} and present example calculations of box slabs excited by monopole and dipole point sources at special frequencies extracted from the dispersion relations. In \S\ref{sec:localized} we demonstrate the application of our approach to localized modes in phononic box arrays.  We conclude in \S\ref{sec:conclude} with a discussion of our results, generalisations to related geometries and other future directions.

\section{Formulation}
\label{sec:formulation}
\subsection{Geometry}
We consider phononic crystals formed of thin-walled perforated boxes arranged on a discrete lattice (see Fig.~\ref{fig:schematic}). The pitch of the lattice is $2a$ and the holes in the perforated faces are circles of radius $\epsilon a$. It is convenient to adopt a dimensionless convention where lengths are normalized by $a$. In particular, let $\bx$ be a normalized position vector measured from the center of an arbitrarily chosen zeroth box. The positions of the box centers are 
\begin{equation}
\bx_c(\vec{n})=2\vec{n}, \quad \vec{n}=n_1\be_1+n_2\be_2+n_3\be_3,
\end{equation}
where $(n_1,n_2,n_3)\in \mathbb{Z}^3$ and $(\be_1,\be_2,\be_3)$ is a right-handed orthogonal basis; note that we use an arrow notation to denote discrete vectors. The corresponding positions of the hole centers are 
 \begin{equation}
 \bx_h(\vec{n},\vec{m}) = \bx_c(\vec{n})+\vec{m}, 
\end{equation}
where $\vec{m}=\pm\be_j$ and $j=1,2$ or $3$. 

The above notation describes the fully three-dimensional arrangement of boxes shown in Fig.~\ref{fig:schematic}(d). We also consider two variants of the phononic box crystal. The first is a quasi-two-dimensional slab of doubly
periodic connected boxes, as shown in Fig.~\ref{fig:schematic}(c); in this case $n_3=0$ for all boxes, and $j=1$ or $2$, since the
top and bottom surfaces do not contain holes. The second is a quasi-one-dimensional array of connected boxes, as shown in Fig.~\ref{fig:schematic}(b); here $n_2=n_3=0$ for all boxes and $j=1$ for all holes. Note that the total number of holes for a single box is twice the dimension $d$ of the array.

It is also convenient to introduce notation for domains and boundaries. Let the overall fluid domain be $B$ with boundary $\partial B$, let $B(\vec{n})$ be the cubic domain of box $\vec{n}$ with boundary $\partial B(\vec{n})$ (approached from inside that box); 
the dimensionless box volume, $\mathcal V$, is $\mathcal{V}=8$ for the cubic boxes we consider but we leave $\mathcal V$ in formulae to allow for ease of generalisation.

\subsection{Acoustic model}
We consider fixed angular frequency, $\omega$, excitation or modes with harmonic time dependence $\exp(-i\omega t)$ assumed understood and henceforth suppressed. We neglect viscous and thermal losses and assume that the box  walls are perfectly rigid. The natural quantity to use is the velocity potential, $\varphi'$, that is related to the physical pressure via $p=i\rho\omega \varphi'$, where $\rho$ is the density of the fluid. In what follows we consider $\varphi$, the potential $\varphi'$ normalised by some reference value. The scaled potential satisfies the forced Helmholtz equation
\begin{equation}
\nabla_{\bx}^2\varphi + \epsilon\Omega^2\varphi = f(\bx), \quad \bx\in B
\label{eq:helmholtz}
\end{equation}
and the Neumann boundary condition
\begin{equation}\label{bc}
\bn\bcdot\bnabla_{\bx}\varphi=0, \quad \bx\in \partial B,
\end{equation}
where $\bn$ is the normal vector pointing into the fluid. In \eqref{eq:helmholtz}, $f(\bx)$ is a possible forcing term and $\Omega$ is a normalised frequency defined via
\begin{equation}\label{subwave}
\frac{a \omega}{c}=\epsilon^{1/2}\Omega. 
\end{equation} 

Our interest is in the limit of small holes $\epsilon\to0$ with $\Omega=O(1)$, i.e., subwavelength frequencies on the order of the resonant frequency of an isolated box. We shall explore this asymptotic limit in the following two cases:

\subsection{Bloch eigenvalue problem}
\label{sec:bloch-eigenvalue}
 The Bloch eigenstates provide the fundamental modes of the periodic system and thereby encapsulate a lot of information about its behaviour. These eigenstates satisfy Eqs.~\eqref{eq:helmholtz} and \eqref{bc} with the forcing set to zero, $f(\bx)\equiv 0$, in conjunction with Floquet--Bloch periodicity,
\begin{equation}\label{quasiperiod}
\varphi(\bx)=\phi(\bx)\exp(i\bb{\kappa}\bcdot \bx), \quad \phi(\bx+2\vec{n})=\phi(\bx),
\end{equation}
where $\bb{\kappa}$ is the Bloch wavevector. The goal is to obtain the eigenfunctions $\varphi$, up to an arbitrary multiplicative constant, together with the dispersion relation $\Omega(\bb{\kappa})$ relating the frequency to the Bloch wavevector. Using \eqref{quasiperiod}, it is possible to formulate the eigenvalue problem over a single unit cell, though this will not be necessary here. 

\subsection{Forced problem}
\label{sec:green}
Identifying the dispersion curves is just a single, all be it important, aspect of modelling crystal media. Equally important is the accurate simulation of forced problems. Without loss of generality, we shall assume that the forcing, $f$, vanishes outside the zeroth box $B(\vec{0})$. In this scenario, a radiation condition is imposed, ensuring outward propagating solutions. Inside the periodic crystal this means that, at large distances from the zeroth box, the solution is composed only of Bloch waves that are outward propagating. 

\section{Discrete lattice model} 
\label{sec:bloch}
\subsection{Outer regions}
\label{ssec:outer}
We begin with the non-forced case, where $f(\bx)\equiv 0$. Consider first the outer limit, $\epsilon\to0$ with $\bx$ fixed, whereby the holes shrink to the points $\bx_h$. Since eigenstates are determined up to a multiplicative constant, we may assume without loss of generality that the outer potential is $O(1)$. In each box $\vec{n}$ we expand the outer potential separately, 
\begin{equation}
\varphi\ub{\vec{n}}(\bx)\sim \varphi_0\ub{\vec{n}}(\bx) + \epsilon \varphi_1\ub{\vec{n}}(\bx) + \cdots, \quad \bx\in B(\vec{n}).
\label{eq:series}
\end{equation}

At leading order, \eqref{eq:helmholtz} gives Laplace's equation in each box,
\begin{equation} \label{lap0}
\nabla_{\bx}^2\varphi_0\ub{\vec{n}} = 0, \quad \bx\in B(\vec{n}),
\end{equation}
whereas \eqref{bc} gives the Neumann conditions
\begin{equation}\label{bc0}
\bn\bcdot\bnabla_{\bx}\varphi_0\ub{\vec{n}} = 0, \quad \bx\in\partial B(\vec{n})/\bx_h(\vec{n},\vec{m})
\end{equation}
on the box faces, possibly excluding the points $\bx_h$. In principle, the behaviour at those points should be determined by matching to the inner regions. In particular, the outer potentials may be singular at the points $\bx_h$, which are excluded from the outer domain. At the present order, however, any such singularity would imply inner potentials that are large in magnitude compared with the outer potentials. This, in turn, implies attenuating inner solutions which, as will become evident, is a contradiction. Thus the Neumann conditions \eqref{bc0} in fact apply uniformly over $\partial B(\vec{n})$. The form of the outer solution is accordingly 
\begin{equation}
\varphi_0\ub{\vec{n}}(\bx) = u\ub{\vec{n}}, \quad \bx\in B(\vec{n}),
\end{equation}
where the $u\ub{\vec{n}}$ are constants. 

Next, the $O(\epsilon)$ balance of \eqref{eq:helmholtz} gives a Poisson equation in each box
\begin{equation}
\nabla_{\bx}^2\varphi_1\ub{\vec{n}} = -{\Omega}^2{u}\ub{\vec{n}}, \quad \bx\in B(\vec{n}),
\label{eq:poisson}
\end{equation}
whereas from \eqref{bc} we again have the Neumann conditions
\begin{equation}\label{bc1}
\bn\bcdot\bnabla_{\bx}\varphi_1\ub{\vec{n}} = 0, \quad \bx\in\partial B(\vec{n})/\bx_h(\vec{n},\vec{m}),
\end{equation}
which as before do not necessarily apply at the points $\bx_h$. Integration of \eqref{eq:poisson} over $B(\vec{n})$, followed by application of the divergence theorem, yields
\begin{equation}\label{gauss}
\underset{\partial{B(\vec{n})}}{\oint}\bn\bcdot\bnabla_{\bx}\varphi_1\ub{\vec{n}}\,d^2\bx = -\mathcal{V}{\Omega}^2{u}\ub{\vec{n}},
\end{equation}
where  the normal $\bn$ points away from ${B}(\vec{n})$. 
To make sense of the left hand side of \eqref{gauss}, recall from our earlier conclusion that the inner potentials are at most $O(1)$. If the scale of the inner region is $\epsilon$, it follows that the singularity of $\varphi_1\ub{\vec{n}}$ as $\bx\to\bx_h$ cannot be stronger than a monopole. As a consequence, the limits
\begin{equation}\label{fluxes}
q\ub{\vec{n},\vec{m}}= -2\pi\lim_{r\to0}\left(r^2\pd{\varphi_1\ub{\vec{n}}}{r}\right), \quad  r=|\bx-\bx_h(\vec{n},\vec{m})|,
\end{equation}
exist and $q\ub{\vec{n},\vec{m}}$ has a natural interpretation as the leading $O(\epsilon)$ flux out of box $\vec{n}$, through the hole centered at $\bx_h(\vec{n},\vec{m})$. Thus \eqref{gauss}, together with \eqref{bc1} and \eqref{fluxes},  yields the connection between the fluxes in, and out, of any given box through its holes, and the uniform potential in that box,
\begin{equation}\label{gauss bloch}
\sum_{\vec{m}} q\ub{\vec{n},\vec{m}}= - \mathcal{V}{\Omega}^2u\ub{\vec{n}}.
\end{equation}
As it stands, the fluxes and potentials are coupled through \eqref{gauss bloch} but remain undetermined. We require a second relation, between the flux through a hole and the potentials in the adjacent boxes, and that is obtained by  turning to an inner problem considering the precise details of the flow in the neighbourhood of that hole.

\subsection{Inner regions}
\label{sec:inner}
Consider next the region localized about the hole centered at $\bx_h(\vec{n},\vec{m})$. The corresponding inner limit is defined as $\epsilon\to0$ with the stretched position vector $\bb{X}$, with 
\begin{equation}
\epsilon\bb{X}=\bx-\bx_h(\vec{n},\vec{m}),
\end{equation}
fixed. We denote the potential in this inner region as $\varphi(\bx)=\Phi(\bb{X})$. The geometry of the inner problem consists of a plane boundary $\bb{X}\bcdot\vec{m}=0$, perforated by a single circular hole of radius unity and centered at $\bb{X}=0$. This boundary is labeled $\partial{W}$ and the unbounded fluid domain surrounding it $W$. 

Under the above inner scaling, the unforced governing equation \eqref{eq:helmholtz} becomes 
\begin{equation}
\nabla^2_{\bb{X}}\Phi+\epsilon^3\Omega^2\Phi=0, \quad \bb{X}\in W,
\label{eq:inner}
\end{equation}
where $\bnabla_{\bb{X}}=\epsilon\bnabla_{\bx}$ is the gradient operator whose Cartesian components are the partial derivative operators with respect to the corresponding components of $\bb{X}$.
Similarly, the Neumann condition \eqref{bc} reads as
\begin{equation}
\bn\bcdot\bnabla_{\bb{X}}\Phi=0, \quad \bb{X}\in \partial W.
\end{equation}
The inner potential is also required to asymptotically match with the outer potentials on both sides of the holes, i.e., in the far-field limits
 $\bb{X}\bcdot\vec{m}\to \pm\infty$. 
 
The outer solution suggests the expansion 
\begin{equation}
\Phi(\bb{X})\sim \Phi_{0}(\bb{X})+\epsilon\Phi_1(\bb{X})+\cdots.
\end{equation}
At leading order \eqref{eq:inner} reduces to Laplace's equation
\begin{equation}\label{inner lap}
\nabla^2_{\bb{X}}\Phi_0=0, \quad \bb{X} \in W,
\end{equation}
while \eqref{bc} gives the Neumann condition
\begin{equation}\label{inner bc}
\bn\bcdot\nabla_{\bb{X}}\Phi_0=0, \quad \bb{X} \in \partial W.
\end{equation}
Turning to the matching, we consider the inner limit of the outer expansion and deduce that on one side of the hole
\begin{equation}\label{inner match right}
\Phi_0 \sim u\ub{\vec{n}+\vec{m}}- \frac{q\ub{\vec{n},\vec{m}}}{2\pi|\bb{X}|} +o\left(\frac{1}{|\bb{X}|}\right) \quad \text{as} \quad \bb{X}\bcdot\vec{m}\to \infty,
\end{equation}
whereas on the other side
\begin{equation}\label{inner match left}
\Phi_0 \sim u\ub{\vec{n}} + \frac{q\ub{\vec{n},\vec{m}}}{2\pi|\bb{X}|} +o\left(\frac{1}{|\bb{X}|}\right) \quad \text{as} \quad \bb{X}\bcdot\vec{m}\to -\infty.
\end{equation}

The solution to the inner problem \eqref{inner lap}--\eqref{inner match left} was known to Lord Rayleigh \cite{rayleigh1870}, see also \cite{howe98a,tuck75a} for more modern treatments. The key point is that, for given flux, the problem is fully determined up to a single additive constant, hence the two uniform potentials either side of the hole, $u\ub{\vec{n}+\vec{m}}$ and $u\ub{\vec{n}}$, are not independent. The flux, $q\ub{\vec{n},\vec{m}}$, is found from the solution to be proportional to their difference 
\begin{equation}
q\ub{\vec{n},\vec{m}}= \beta\left(u\ub{\vec{n}+\vec{m}}-u\ub{\vec{n}}\right),
\label{eq:connect}
\end{equation}
where, for the circular hole we have chosen the constant of proportionality, $\beta=2$; other aperture geometries, including the possibility of a wall thickness commensurate with the hole radius, yield a similar result only with different values for $\beta$ \cite{howe98a,tuck75a}. 

\subsection{Asymptotic discrete model and dispersion relation} 
Combining \eqref{eq:connect} with \eqref{gauss bloch}, we arrive at the discrete wave equation
\begin{equation}
\sum_{\vec{m}}u\ub{\vec{n}+\vec{m}}+\left( \frac{\mathcal{V}}{\beta}{\Omega}^2-2d\right)u\ub{\vec{n}}=0,
\label{eq:result}
\end{equation}
where, for each box $\vec{n}$, the sum is taken over the 
  vectors $\vec{m}$ associated with that box and $2d$ is the number of holes in each box. Thus far we have not enforced Bloch--Floquet conditions, and \eqref{eq:result} remains a general model of unforced phononic box crystals. We now specialise to Bloch waves, in which case \eqref{quasiperiod} implies the ansatz
\begin{equation}\label{ansatz}
u\ub{\vec{n}}= \text{const.} \times \exp(i\bb{\kappa}\bcdot \bx\ub{\vec{n}}).
\end{equation}
For the triply periodic phononic box crystal ($d=3$), substituting \eqref{ansatz} into \eqref{eq:result} yields the dispersion relation
\begin{equation}
{\Omega}^2=\frac{2\beta}{\mathcal{V}}\left[3-\cos(2\kappa_x)-\cos(2\kappa_y)-\cos(2\kappa_z)\right]. 
\label{eq:cubic}
\end{equation}
The dispersion relations for quasi-2D ($d=2$) and quasi-1D ($d=1$) phononic box arrays are obtained simply by setting the corresponding components of the  Bloch wavevector to zero.
 
\begin{figure}[t!]
\begin{center}
\includegraphics[scale=0.5]{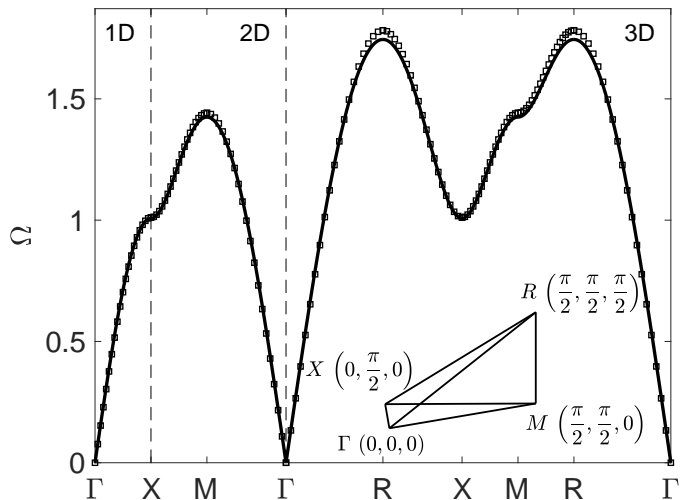}
\caption{Dispersion curves for hole radius $\epsilon=10^{-1}$ showing both finite element (symbols) and asymptotic (solid line) results. In the finite element simulations the wall thickness is finite, but small compared to both the pitch and the hole radius. 
The curves for the quasi-1D and quasi-2D arrangements are a subset of the full 3D ones. The inset shows the irreducible Brillouin zone.}  
\label{fig:dispersion}
\end{center}
\end{figure}

In Fig.~\ref{fig:dispersion} we show, for a relatively large value of $\epsilon=0.1$, the dispersion for both the lattice model \eqref{eq:cubic} and, for comparison, finite element computations \cite{comsol}. Recalling that our discrete asymptotic model was developed for $\epsilon\ll 1$, the agreement is pleasing. The inset to Fig.~\ref{fig:dispersion} shows the irreducible Brillouin zone \cite{brillouin53a} and, as is conventional in the physics literature, we plot the dispersion curves going around the edges of this zone connecting the high symmetry points. Note that the curves for the quasi-2D and quasi-1D cases are subsets of these curves. It is clear that both qualitative and quantitative details are captured by the asymptotic method and therefore features of physical interest such as frequencies where dynamic anisotropy is observed, or the effective refractive index in the long-wavelength limit, can be extracted. 

\subsection{Effective index}
Recalling the scaling \eqref{subwave}, the dispersion relation \eqref{eq:result} predicts nonlinear dispersion in a wide subwavelength regime, where the wavelength in free space is $O(\epsilon^{-1/2})$ compared to the box dimensions. At the lower end of this subwavelength regime, i.e., for $\Omega\ll1$, it is evident from \eqref{eq:result}, and from the example shown in Fig.~\ref{fig:dispersion}, that the dispersion becomes linear. In that limit, the dispersion is simply captured by an effective index of refraction, which is readily extracted from the small $\bb{\kappa}$ limit of \eqref{eq:result}, $\Omega^2 \sim (4\beta/\mathcal{V})|\bb{\kappa}|^2$. Together with \eqref{subwave} we obtain the large effective index 
\begin{equation}
n_{\text{eff}}\sim \left(\frac{\mathcal{V}}{4\epsilon\beta}\right)^{1/2}=\epsilon^{-1/2} \qquad (\Omega\ll1). 
\end{equation}

\section{Waves induced by a distribution of sources} 
\label{sec:forced}
When using the asymptotic approach to generate the solutions induced by sources, it is particularly important to note that the form of the source can lead to differences in the asymptotic scaling of the solution. It is tempting to simply introduce a source-like term directly into the discrete network model \eqref{eq:result}, but the form of such a term needs to be derived so one can go precisely from the continuum level to the discrete system. 
\subsection{Monopole point source}
We begin by considering the problem where the forcing consists of a monopole point source, i.e., $f(\bx)=\delta(\bx-\bx_s)$, assuming that $\bx_s\in B(\vec{0})$ excluding the $O(\epsilon)$ vicinities of the holes. 

In free space, i.e., in the absence of any walls, the above forcing would obviously yield an $O(1)$ response (namely the free-space Green's function). Let us begin by showing that naively assuming the same scaling  for the phononic box crystal leads to a contradiction. Indeed, posing an outer expansion in the form \eqref{eq:series}, we find at leading order the Poisson equation 
\begin{equation}\label{false poisson}
\nabla_{\bx}^2\varphi\ub{\vec{n}}_{0}(\bx,\bx_s)=\delta(\bx-\bx_s), \quad \bx\in B(\vec{n}),
\end{equation}
instead of Laplace's equation \eqref{lap0}. 
Clearly the delta forcing is incompatible with the Neumann condition \eqref{bc0}, which for the same reasons as in \S\S\ref{ssec:outer} would appear to hold over the entire boundary $\partial{B}(\vec{0})$. The escape from this conundrum is to recognise that the expansion \eqref{eq:series} is incorrect and $O(\epsilon^{-1})$ potentials are needed in order to accommodate the $O(1)$ flux generated by the delta forcing. 

In light of the above, we pose the modified outer expansion
\begin{equation}
\varphi\ub{\vec{n}}(\bx)\sim \epsilon^{-1}\varphi\ub{\vec{n}}_{-1}(\bx)+\varphi\ub{\vec{n}}_{0}(\bx)+\cdots, \quad \bx\in B(\vec{n}).
\end{equation}
A leading-order analysis, similar to before, shows that
\begin{equation}
\varphi\ub{\vec{n}}_{-1}(\bx) =v\ub{\vec{n}}, \quad \bx\in B(\vec{n})
\end{equation}
 where the $v\ub{\vec{n}}$ are constants. At the next order, we obtain
\begin{equation}\label{forced poisson}
\nabla_{\bx}^2\varphi\ub{\vec{n}}_{0} = -{\Omega}^2{v}\ub{\vec{n}} + \delta(\bx-\bx_s), \quad \bx\in B(\vec{n}),
\end{equation}
together with the Neumann conditions 
\begin{equation}
\bn\bcdot\bnabla_{\bx}\varphi\ub{\vec{n}}_{0} = 0, \quad \bx\in \partial B(\vec{n})/\bx_h(\vec{n},\vec{m}).
\end{equation}
For all but the zeroth box, the Poisson equation \eqref{forced poisson} reduces to \eqref{eq:poisson} (except for the ordering offset) and the network equations are simply obtained from the non-forced network \eqref{eq:result} by substituting $u\ub{\vec{n}}$ with the enhanced field  $v\ub{\vec{n}}$. As for the zeroth box, the analysis proceeds similarly to the non-forced case, only that when integrating \eqref{forced poisson} over $B(\vec{0})$ the Dirac delta forcing produces an extra inhomogeneous constant term of unit value. We accordingly arrive at the following forced network model:
\begin{equation}
\sum_{\vec{m}}v\ub{\vec{n}+\vec{m}}+\left( \frac{\mathcal{V}}{\beta}{\Omega}^2-2d\right)v\ub{\vec{n}}=\frac{1}{\beta}\delta_{\vec{n},\vec{0}},
\label{eq:forced_discrete}
\end{equation}
where here $\delta$ is the Kronecker delta function. 

\begin{figure}[t!]
  \centering
\includegraphics[scale=0.32]{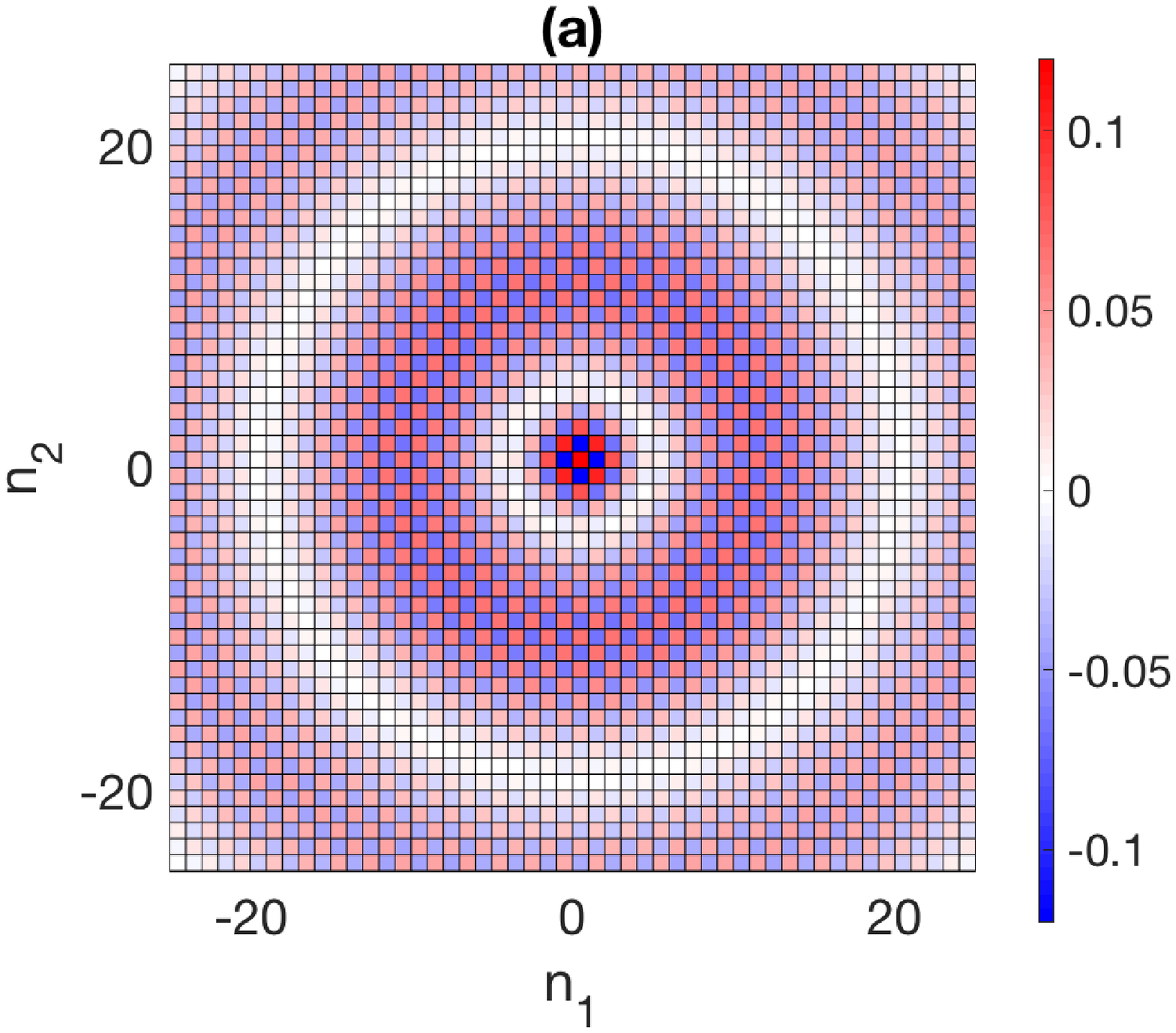}
\includegraphics[scale=0.32]{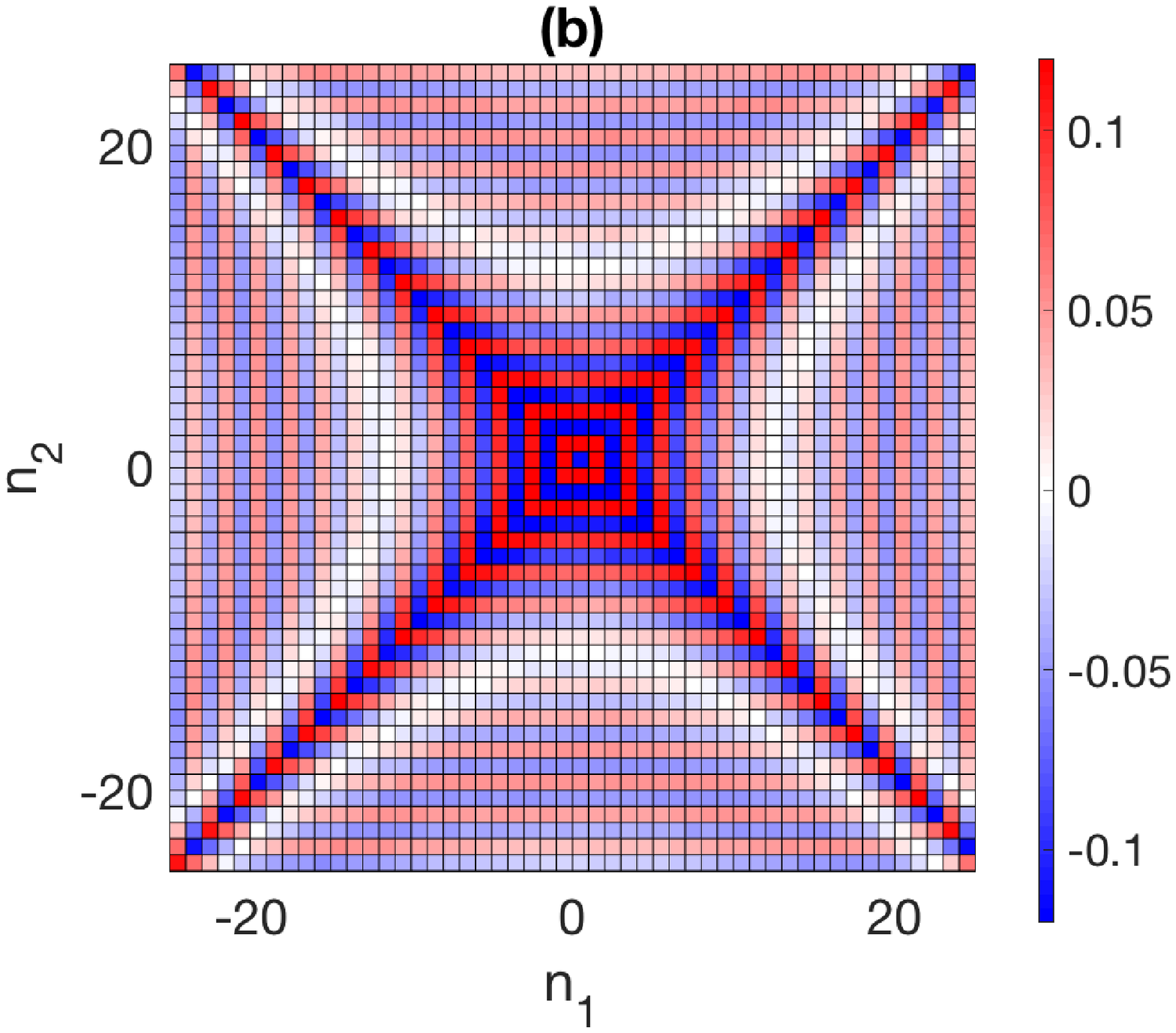}
  \caption{The enhanced field $v\ub{\vec{n}}$ generated by a monopole source for a quasi-2D slab of boxes with the forcing in the zeroth box. Panels (a) and (b) show the real part of the potential for frequencies $\Omega=1.4107$ and $0.995$, respectively.}
\label{fig:monopole_forced}
\end{figure}

The discrete lattice model \eqref{eq:forced_discrete} is easily solved numerically, allowing one to simulate large-scale box crystals that would be computationally demanding to simulate otherwise. In Fig.~\ref{fig:monopole_forced} we show illustrative simulations in the case of a quasi-2D slab of boxes, performed using the methodology from \cite{vanel16}. These examples demonstrate the strongly frequency-dependent nature of the fields; referring back to Fig.~\ref{fig:dispersion}, we choose frequencies close to the high-symmetry points $\textrm{X}$ and $\textrm{M}$ and observe that the monopole forcing can have either a highly isotropic response, Fig.~\ref{fig:monopole_forced}(a), or highly directional anisotropy, Fig.~\ref{fig:monopole_forced}(b). 

\subsection{General source distribution}\label{ssec:general}
An interesting feature of the above leading-order solution is its independence upon the precise position $\bx_s$ of the monopole source. This point is very useful for, say, an arbitrary distribution of sources in the zeroth box, in which case superposition readily gives
\begin{equation}\label{general source}
\varphi(\bx) \sim  \epsilon^{-1}Fv\ub{\vec{n}} +O(1), \quad \bx\in B(\vec{n})
\end{equation}
where $F$ is the total monopole strength of the forcing, i.e., the integral of the source distribution $f(\bx)$ over the zeroth box. Thus, in particular, the $O(1/\epsilon)$ field enhancement remains valid for an arbitrary distribution of sources provided that $F\neq 0$. 

If, however, $F=0$, the enhanced field vanishes and \eqref{general source} fails to provide the leading-order excitation. There are two ways to address this special case. The first is to continue the analysis in the case of a monopole point source to one higher order and again use superposition. This would give not only the leading-order approximation for a source distribution with $F=0$, but also an improved two-term approximation in the case where $F\ne0$. An alternative, more efficient and illuminating approach, is to consider the case $F=0$ independently. Accordingly, in what follows we develop the leading-order approximation for the excitation due to a single \emph{dipole} point source, whose orientation and position in the zeroth box are arbitrary. From this problem, the general case of an arbitrary continuous distribution of dipoles can again be found by superposition although for brevity we omit the details herein.

\subsection{Dipole point source}\label{ssec:dipole}
Thus consider the forcing $f(\bx)=\bk\bcdot\bnabla\delta(\bx-\bx_s)$, where $\bk$ is a unit vector and as before $\bx_s$ is within the zeroth box, not too close to the holes. For a dipole forcing the incompatibility encountered in \S\S\ref{ssec:outer} when assuming an $O(1)$ field does not arise and, thus, in accordance with the remarks made in \S\S\ref{ssec:general}, there is no asymptotic enhancement of the field in the case of a dipole source. 

We accordingly expand the outer potential in box $\vec{n}$ as
\begin{equation}
\varphi\ub{\vec{n}}(\bx;\bk)\sim \varphi\ub{\vec{n}}_{0}(\bx;\bk)+\epsilon\varphi\ub{\vec{n}}_{1}(\bx;\bk)+\cdots, \quad \bx\in B(\vec{n}).
\end{equation}
At leading order we find the Poisson equations 
\begin{equation}\label{Poisson dipole}
\nabla_{\bx}^2\varphi\ub{\vec{n}}_0 = \bk\bcdot\bnabla_{\bx}\delta(\bx-\bx_s), \quad \bx\in B(\vec{n})
\end{equation}
and Neumann boundary conditions
\begin{equation}\label{bc0 dipole}
\bn\bcdot\bnabla_{\bx}\varphi\ub{\vec{n}}_0=0, \quad \bx\in\partial B(\vec{n}),
\end{equation}
where the latter hold over the entire boundaries $\partial B(\vec{n})$, without need for singularities at this order. For all but the zeroth box, we find as usual that
\begin{equation}
\varphi\ub{\vec{n}}_0=u\ub{\vec{n}}, \quad \bx\in B(\vec{n})/ B(\vec{0}),
\end{equation}
where the constants $u\ub{\vec{n}}$ remain to be determined. 

Consider now the zeroth box, where the forcing term in \eqref{Poisson dipole} clearly generates a non-uniform leading-order potential. To develop the solution we introduce a Neumann Green's function $G(\bx,\bx')$ for the zeroth box, solving
\begin{equation}\label{G eq}
\nabla_{\bx}^2G = -\delta(\bx-\bx')+\frac{1}{\mathcal{V}}, \quad \bx,\bx'\in B(\vec{0}),
\end{equation}
together with a Neumann boundary condition
\begin{equation}\label{G bc}
\bn\bcdot\bnabla_{\bx}G=0, \quad \bx \in \partial B(\vec{0}).
\end{equation}
Note that only the gradient of the function $G$ is uniquely determined.  Applying Green's second identity to the pair $G$ and $\varphi\ub{\vec{0}}_0$, using the Poisson equations \eqref{Poisson dipole} and \eqref{G eq} and the Neumann conditions \eqref{bc0 dipole} and \eqref{G bc}, we find the following integral representation of the potential in the zeroth box:
\begin{equation}
\varphi\ub{\vec{0}}_0(\bx') = u\ub{\vec{0}} -\int_{B(\vec{0})} G(\bx,\bx')\bk\bcdot\bnabla_{\bx}\delta(\bx-\bx_s)\,d^3\bx,
\label{eq:Dzero}
\end{equation}
where $u\ub{\vec{0}}$ is defined as the average
\begin{equation}
u\ub{\vec{0}} =\frac{1}{\mathcal{V}}\int_{B(\vec{0})} \varphi\ub{\vec{0}}_0(\bx)\,d^3\bx.
\end{equation}
Noting that $\bnabla_{\bx}\delta(\bx-\bx_s)=-\bnabla_{\bx_s}\delta(\bx-\bx_s)$, the operator $\bk\bcdot\bnabla_{\bx_s}$ can be taken out of the integral, whereby the sifting property of the Dirac delta function yields
\begin{equation}\label{general solution}
\varphi\ub{\vec{0}}_0(\bx') = u\ub{\vec{0}} + \bk \bcdot \bnabla_{\bx_s}G(\bx_s,\bx'). 
\end{equation}
Thus the deviation of the potential in the zeroth box from its mean value, which remains to be determined, is given  explicitly in terms of the orientation $\bk$ and location $\bx_s$ of the dipole source. This, of course, assumes that we can calculate the gradient of the Neumann Green's function, a matter which we shall return to shortly.

\begin{figure}[t!]
  \centering
\includegraphics[scale=0.32]{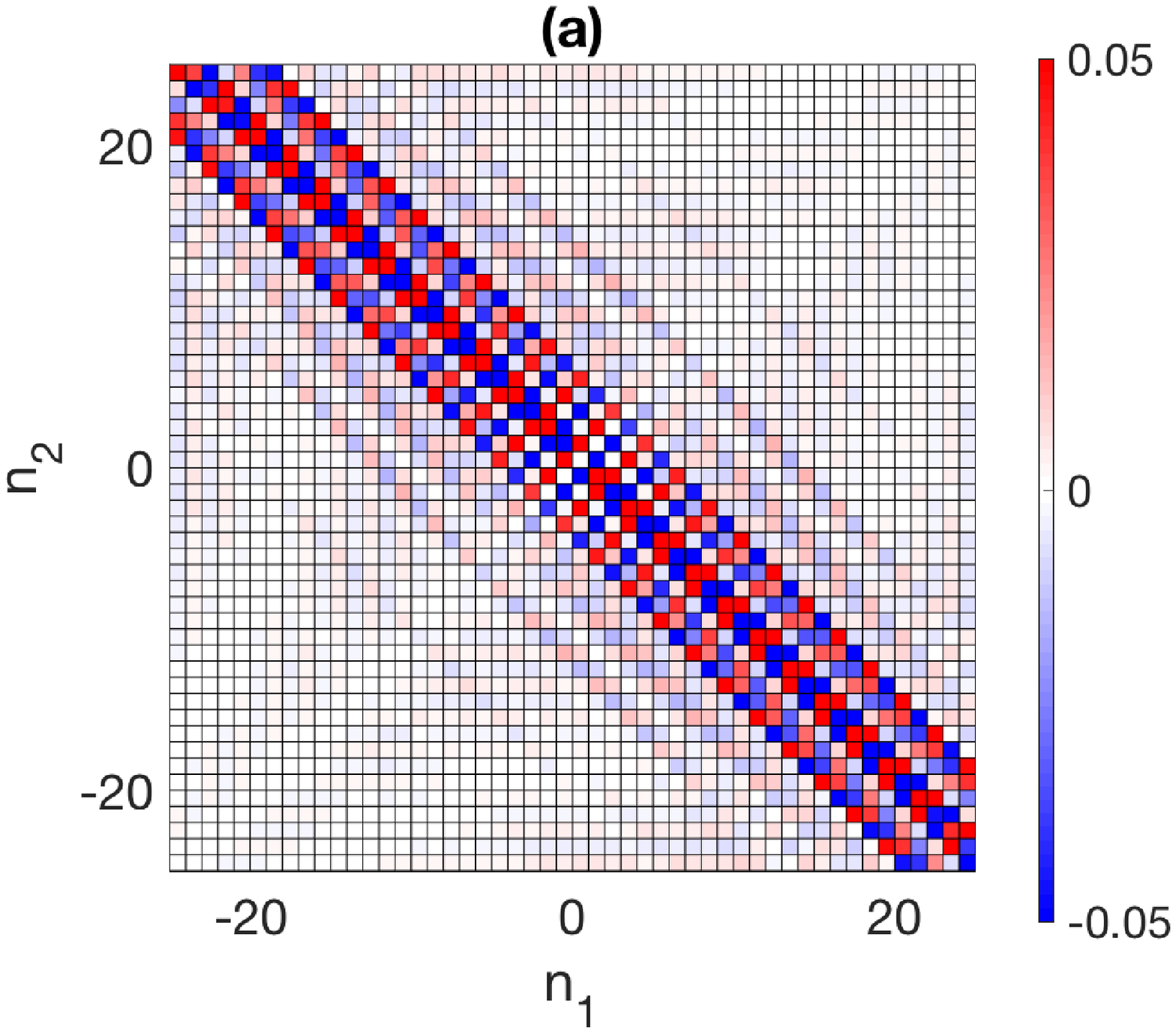}
\includegraphics[scale=0.32]{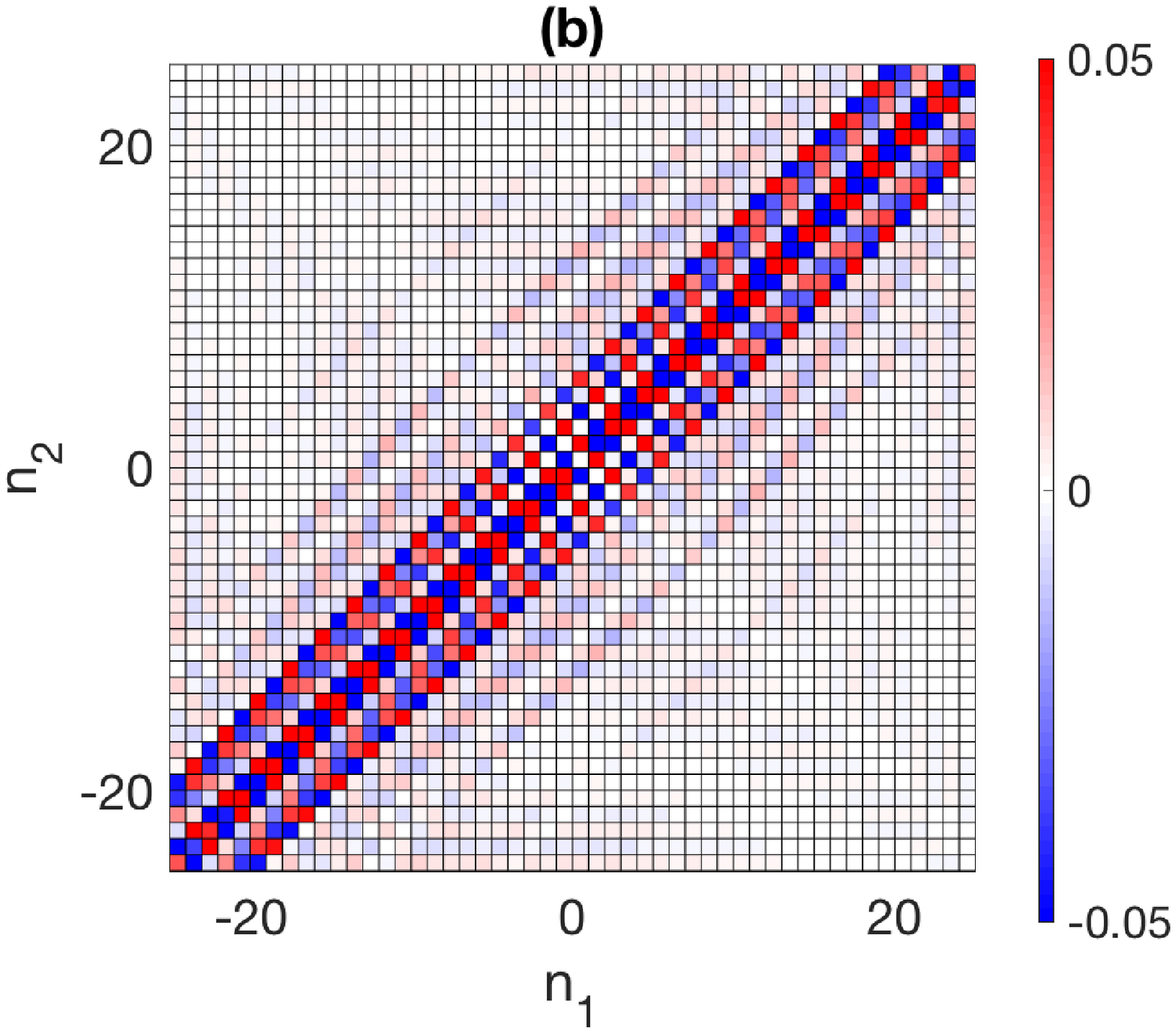}
\includegraphics[scale=0.32]{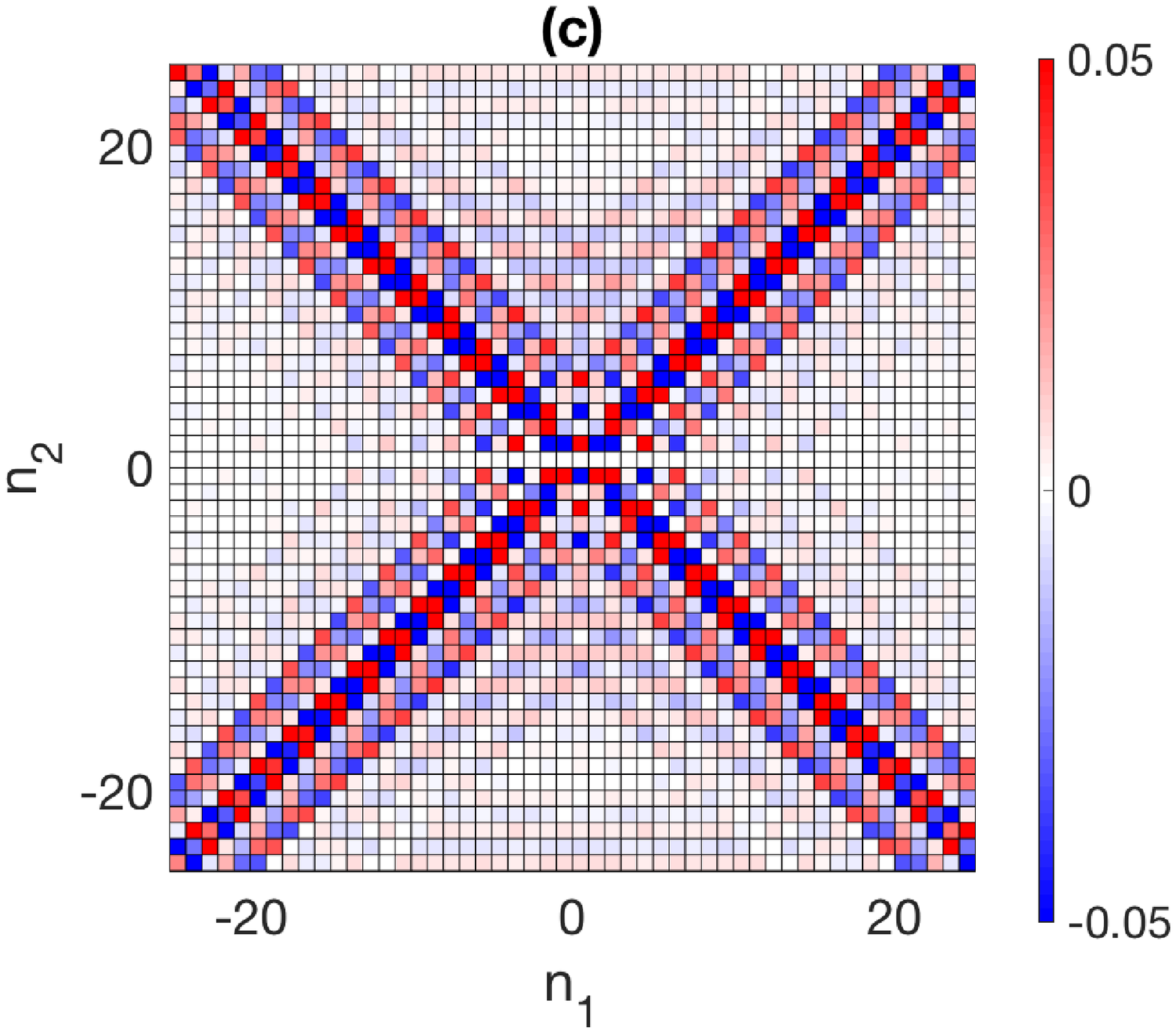}
\includegraphics[scale=0.32]{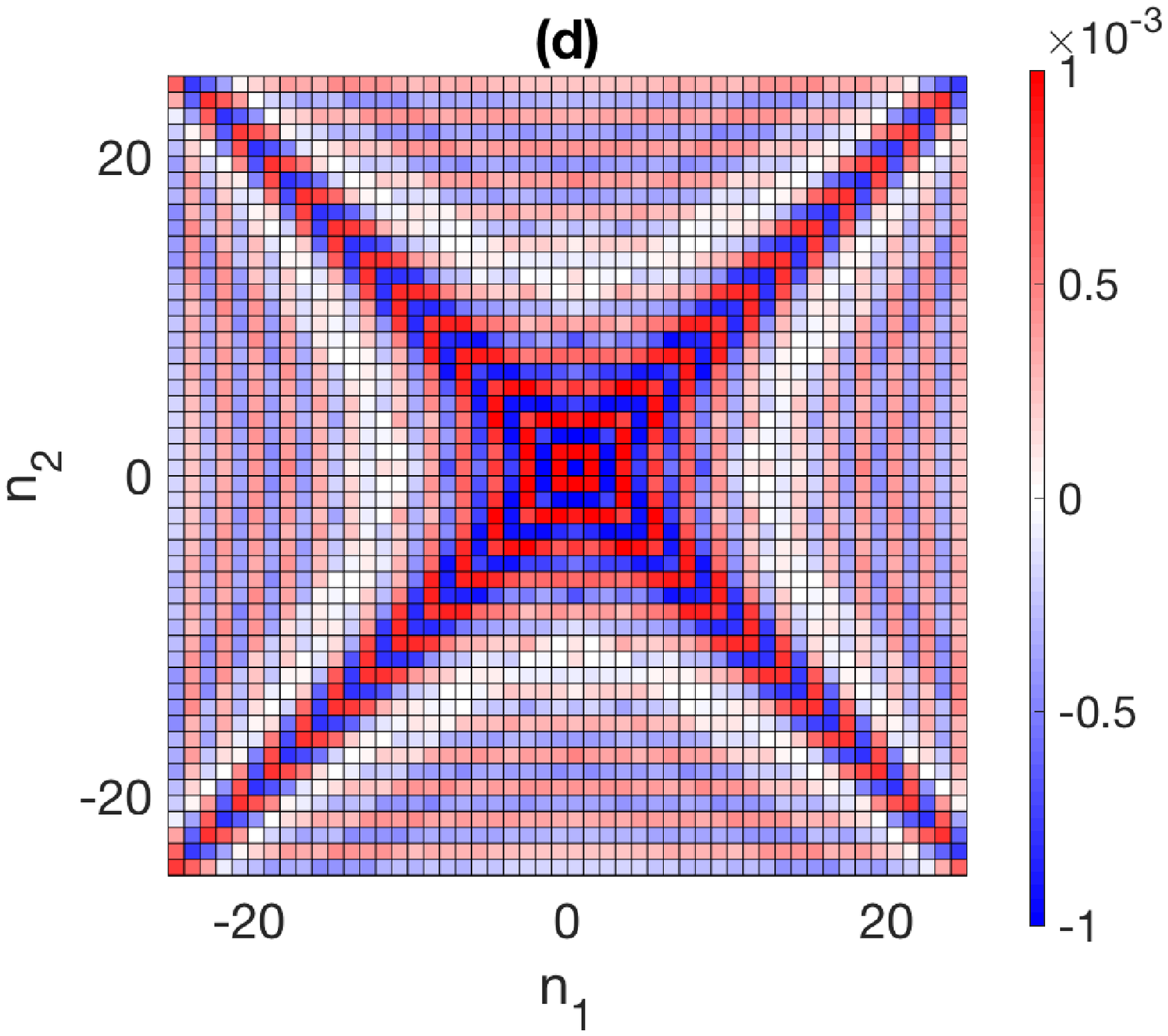}
  \caption{
Dipolar forcing at frequency $\Omega=0.995$, for the quasi-2D slab of boxes with the dipole in the zeroth box at $\bx_s=(0,0,0)$. The real part of the field is shown for four orientations: (a) $\bk=(1/\sqrt{2},1/\sqrt{2},0)$, (b) $\bk=(-1/\sqrt{2},1/\sqrt{2},0)$ (c) $\bk=(1,0,0)$ and (d) $\bk=(0,0,1)$. This frequency gives highly anisotropic responses and the direction of the dipole determines the energy propagation direction. The pressure is uniform in each box except the zeroth where we show the average.}
\label{fig:dipole_forced_X}
\end{figure}

At the next order we have the Poisson equation
\begin{equation}
\nabla_{\bx}^2\varphi\ub{\vec{n}}_1 = -{\Omega}^2\varphi\ub{\vec{n}}_0, \quad \bx\in B(\vec{n})
\end{equation}
 subject to the Neumann conditions
\begin{equation}
\bn\bcdot\bnabla_{\bx}\varphi\ub{\vec{n}}_1=0, \quad \bx\in \partial B(\vec{n})/\bx_h(\vec{n},\vec{m})
\end{equation}
and matching conditions at the singular points $\bx_h(\vec{n},\vec{m})$. Integrating over the volume of the $\vec{n}$th box again leads to \eqref{gauss bloch}, with $q\ub{\vec{n},\vec{m}}$ defined just as in \eqref{fluxes}. Away from the zeroth box, the canonical inner problems arise as in section \ref{sec:inner} and we again obtain \eqref{eq:connect}. The zeroth box introduces some novelty, however, as the relevant potential affecting the flux through any of its holes is not the mean $u\ub{\vec{0}}$, but the limit value of $\varphi\ub{\vec{0}}_0$ as the position of that hole is approached. Concretely, when considering matching with an inner region corresponding to the hole of the zeroth box centered at $\bx=\vec{m}$, the matching condition \eqref{inner match left} becomes
\begin{equation}\label{inner match modified left}
\Phi_0 \sim \lim_{\bx\to \vec{m}} \varphi\ub{\vec{0}}_0(\bx) + \frac{q\ub{\vec{0},\vec{m}}}{2\pi|\bb{X}|} +o\left(\frac{1}{|\bb{X}|}\right) \quad \text{as} \quad \bb{X}\bcdot\vec{m}\to -\infty.
\end{equation}
Defining 
\begin{equation}
\bb{L}\ub{\vec{m}}(\bx_s)= \lim_{\bx'\to\vec{m}} \bnabla_{\bx_s}G(\bx_s,\bx'),
\end{equation}
the general solution \eqref{general solution} gives the limit in \eqref{inner match modified left} as 
\begin{equation}
\lim_{\bx\to \vec{m}} \varphi\ub{\vec{0}}_0(\bx)=u\ub{\vec{0}} +  \bk\bcdot \bb{L}\ub{\vec{m}}(\bx_s).
\end{equation}
Note that calculating the limits $\bb{L}\ub{\vec{m}}(\bx_s)$ entails taking the source position in the Neumann Green's function up to boundary points. In the appendix we develop a regularized and efficient scheme to evaluate these limits. Once the $\bb{L}\ub{\vec{m}}(\bx_s)$ have been calculated, we follow the derivation of \eqref{eq:connect} to obtain a generalized relation between flux and potential differences,
\begin{equation}\label{dipole flux condition}
q\ub{\vec{n},\vec{m}} = \beta\left(u\ub{\vec{n}+\vec{m}}-u\ub{\vec{n}}\right)+\beta\bk\bcdot\left(\delta_{\vec{n}+\vec{m},\vec{0}}\bb{L}\ub{-\vec{m}}(\bx_s)-\delta_{\vec{n},\vec{0}} \bb{L}\ub{\vec{m}}(\bx_s)\right),
\end{equation}
which is written such that it holds for all boxes and holes. Together with \eqref{gauss bloch}, we thus find the discrete system governing the excitation due to a dipole source,
\begin{equation}\label{dipole network}
\sum_{\vec{m}} u\ub{\vec{n}+\vec{m}}+ \left(\frac{\mathcal{V}}{\beta}{\Omega}^2-2d\right)u\ub{\vec{n}} =\bk\bcdot\sum_{\vec{m}} \left[\left(-\delta_{\vec{n}+\vec{m},\vec{0}}\bb{L}\ub{-\vec{m}}(\bx_s)+\delta_{\vec{n},\vec{0}} \bb{L}\ub{\vec{m}}(\bx_s)\right)\right].
\end{equation} 
The forcing terms in \eqref{dipole network} are determined by the non-uniform deviation of the potential in the zeroth box from its mean. As a consequence, and in contrast to the case of a monopole source, the forcing terms depend on the source position (as well as orientation). Moreover, the forcing terms affect not only the zeroth node of the network, but also the surrounding ones. 

In Fig.~\ref{fig:dipole_forced_X} we choose to illustrate the variation in the field for a single dipole placed at the center of the zeroth box in the two-dimensional box slab of Fig.~\ref{fig:schematic}(c). Fig.~\ref{fig:dipole_forced_X}(a-c) show that orientating the dipole across the box or along its $x$ axis can create distinctive dynamic anisotropy whereas orientating the dipole vertically, Fig.~\ref{fig:dipole_forced_X}(d), yields a very similar field to that of the monopole, Fig.~\ref{fig:monopole_forced}(b), but one order of magnitude lower 
than in the other panels of Fig. \ref{fig:dipole_forced_X}.

\section{Localized modes} \label{sec:localized}
As a final example of the utility and general nature of our asymptotic approach, let us briefly consider the possibility of localized modes in defected box arrays, i.e., eigenmodes localized in space and exponentially decaying in the far field,  in a line array of connected boxes with one or more boxes altered. In particular, consider the waveguide configuration shown in Fig.~\ref{fig:schematic}(b), where now the volume of the zeroth box is decreased by insertion of a cubical obstacle. In our asymptotic framework, such a defect simply enters Eq.~\eqref{gauss bloch} by varying the volume of the zeroth box from $\mathcal V$ to $\mathcal V^*$, say. In that way we readily obtain the network model
\begin{equation}
\beta(u_{n+1}-2u_n+u_{n-1})=-\Omega^2u_n
\begin{cases}
\mathcal{V}\qquad &n=0 \\
\mathcal{V^{*}} \qquad & n\ne 0,
\end{cases}
\label{eq:defect}
\end{equation}
where $u_n=u\ub{\vec{n}}$ with $\vec{n}=n\be_1$.

\begin{figure}[b!]
  \centering
\includegraphics[trim={0 0 0 4cm},clip,scale=0.62]{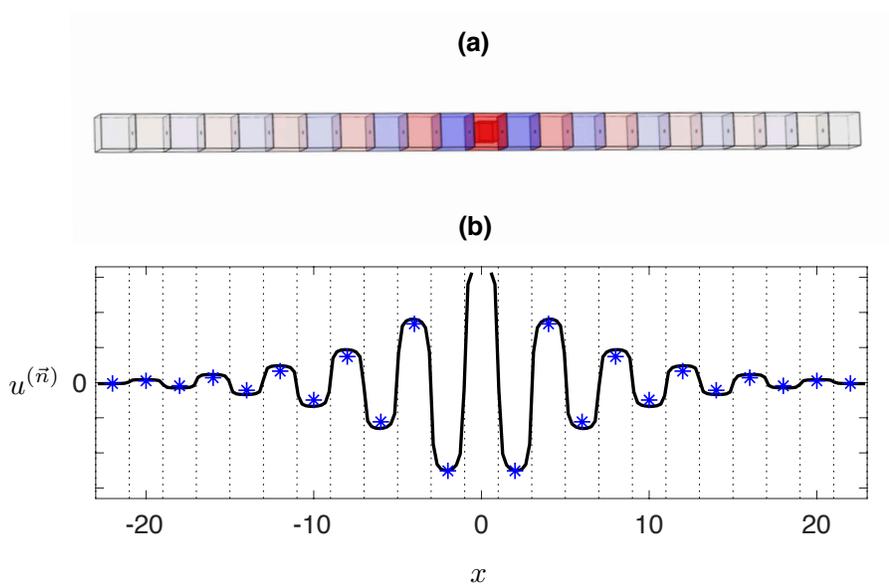}
  \caption{Localized mode in a line waveguide configuration of boxes for holes of radii $\epsilon=0.1$ and finite wall thickness $H/a=10^{-2 }$ where the volume has decreased in the zeroth box. Numerical simulations give the frequency in the bandgap as
 $\Omega\approx1.027$ whereas the asymptotics yield $\Omega\approx1.021$. 
 (a) shows the finite element simulation with the pressure virtually uniform in each box. (b) compares the real part of the field along the centerline, blue stars from the difference equation and solid lines from the simulation. 
}
\label{fig:localized_defect_2}
\end{figure}
The above discrete model is familiar from solid state physics and its eigensolutions are found explicitly using Fourier series as in \cite{bacon62a}. The results yield a single localized mode at an isolated frequency in the bandgap; a typical result,   in Fig.~\ref{fig:localized_defect_2}, shows the numerical simulation from Comsol Multiphysics \cite{comsol} versus the result from solving the difference equations \eqref{eq:defect}. In this example the volume of the zeroth box is reduced from $\mathcal{V}=8$ to $\mathcal{V^{*}}=6.4$. The eigen-frequency obtained from the difference equations is exactly $\Omega^2=8(2\mu-\mu^2)^{-1}\mathcal{V}^{-1}$, where $\mu=\mathcal{V}^{*}/\mathcal{V}$ \cite{bacon62a}; for the present example this yields $\Omega\approx1.021$, in excellent agreement with the result from the finite-element simulations, $\Omega\approx1.027$. Fig.~\ref{fig:localized_defect_2}(b) is instructive showing clearly the nearly constant pressure in each box, and the rapid variation across each hole, together with the predictions of the network model.

\section{Concluding remarks} 
\label{sec:conclude}
Phononic box crystals constitute a family of non-standard metamaterials, in which strongly coupled Helmholtz resonators act in concert to uniformly squeeze the acoustic branch into a wide subwavelength regime. 
Using matched asymptotic expansions, we develop discrete wave equations that accurately model such media in the limit where the holes connecting the boxes are small and the frequency is in the above-mentioned resonant regime. 
 Of course, network equations are ubiquitous in many areas of physics and engineering, for example in solid state physics \cite{kittel96a}. We emphasize that the present ones were systematically derived starting from a continuum description, with the lumped parameters, obtained in closed form, necessarily yielding characteristic frequencies in the subwavelength regime. The physical interpretation of phononic box crystals in terms of their equivalent mass-spring networks is mostly similar to the classical interpretation of an isolated Helmholtz resonator, i.e., the fluid in the vicinity of each hole acts as a mass and each void acts as a spring. Here, however,  the motion of the mass represents fluid flow between two coupled resonators, while the spring force is proportional to the \emph{sum} of mass displacements, i.e., to the net fluid flux through the box holes.

We have used the network equations to derive dispersion relations for Bloch waves propagating in phononic box arrays, slabs and crystals, deducing and demonstrating various properties such as dynamic anisotropy and a high effective index in the long-wavelength limit. We also considered excitation of acoustic waves by sources placed within one of the boxes, finding that there are two fundamentally different classes of forced problems. If the source distribution integrates to give a net monopole, then the pressure is enhanced by a factor of $O(\epsilon^{-1})$; in that case, the pressure remains approximately uniform in the forced box, just as in non-forced boxes, and the forcing simply enters the discrete wave equation as a delta function. It is possible to explain this enhancement effect based on the smallness of the forced box relative to the wavelength, as this implies that the pulsating volume of fluid concomitant with the source distribution cannot be accommodated locally through compression of the fluid; rather, that fluid must be expelled through the tiny holes. This, in turn, necessitates large pressure differences between the boxes. The second class of forced problems arises for dipolar source distributions, for which the net monopole of the source distribution in the forced box vanishes. In this case, there is no asymptotic enhancement and the field in the forced box has a complicated form. We showed, for forcing by a dipole point source, that the appropriate forcing at the discrete level is determined by calculating certain singular limits of the Neumann Green's function for a cube. Using symmetry arguments and asymptotic analysis, we developed a regularized and efficient scheme for calculating these limits. We note that our introduction here of a modified, Neumann, Green's function is typical of analyzes where there are small defects on an otherwise perfect closed Neumann surface, see for instance \cite{coombs09a} for an application in diffusion. 

Our analysis illustrates the power of using matched asymptotic expansions to model resonant metamaterials and 
we conclude by mentioning several extensions specifically for phononic box crystals. First, our asymptotic approach could be adapted to more general box configurations, similarly to the way we considered arbitrary inclusion and lattice geometries in the two-dimensional case \cite{vanel17a}. Indeed, we have already demonstrated this, to some extent, in the brief example given in \S\ref{sec:localized} of a localized mode, where the non-trivial geometry of the defected box posed no special difficulty. It would be similarly straightforward to consider complex networks composed of a variety of coupled resonators, not necessarily cubic, arranged in some arbitrary ordered or even disordered manner. In particular, following \cite{vanel17a}, one could conceive periodic arrangements with unit cells having more than one degree of freedom (on the discrete level), say using multiple volumes or holes. In that case one expects multiple subwavelength branches and thus more complicated dispersion with features such as Dirac points. It would also be possible to incorporate into the model a non-zero wall thickness (say, comparable with the hole dimension), as well as non-circular apertures, simply by resolving, possibly numerically, the appropriate inner potential-flow problem governing the parameter $\beta$.

It would also be desirable to extend our asymptotic approach to hold under more realistic physical conditions. One obvious goal would be to rigorously incorporate viscous and thermal losses on the discrete level, preferably starting from a first-principles description \cite{Pierce:Book}. Depending on the acoustic medium considered, as well as the scale of the box and the holes, which together set the frequency regime of interest, losses may have a significant effect especially owing to the smallness of the holes \cite{Ward:15}. We finally propose to consider phononic box crystals at higher frequencies than considered here. Finite-element simulations of the Bloch eigenvalue problem reveal that above the subwavelength acoustic branch there is a wide band gap, a feature which is highly desirable in many applications for vibration control, such as in seismic metamaterials \cite{achaoui17a,miniaci16a}. Often band-gaps are broadened by inserting multiple isolated resonators, with nearby resonant frequencies \cite{jimenez17a}, but here the effect is obtained utilising the connection between them. Above the band gap, the wavelength is commensurate with the cell size. In that regime, we anticipate that phononic box crystals act more as traditional phononic crystals, however with wave propagation dominated by the cavity resonances of the boxes, rather than multiple-scattering, and with the small gaps still dominating the coupling between the boxes. This raises the question whether this intriguing regime could also be analyzed using asymptotic techniques.  

\section*{Acknowledgements}
The authors thank the EPSRC for their support, RVC and AV via grant number EP/L024926/1 and OS via EP/R041458/1.   Additionally RVC gratefully acknowledges support of a Leverhulme Trust Fellowship. 

\appendix
\section{Limit of Neumann Green's function}
The problem governing the Neumann Green's function $G(\bx,\bx')$, where $\bx,\bx'\in B(\vec{0})$, is described in  \S\S\ref{ssec:dipole} [cf.~\eqref{G eq} and \eqref{G bc}]. Our goal here is to derive a convenient scheme for calculating the limit
\begin{equation}
\underset{\bx'\to\vec{m}}{\lim}\bnabla_{\bx}G(\bx,\bx'),
\end{equation}
where $\vec{m}=\be_l$, with $l=1,2$ or $3$; the position of the source limits to a point on the boundary of the domain. Our approach is to introduce a small positive auxiliary parameter $\lambda$ and to consider the limit,
\begin{equation}\label{lambda def}
\bx'=(1-\lambda)\vec{m}, \quad \lambda\to0.
\end{equation} 
Note that, for small finite $\lambda$, the closeness of the source to the boundary introduces a small length scale relative to the box size. We accordingly treat this limit using matched asymptotic expansions, where the outer limit coincides with the original one, namely $\lambda\to0$ with $\bx$ fixed, whereas the inner limit is $\lambda\to0$ with the stretched coordinate 
\begin{equation}
\bb{Y}=\frac{\bx-\vec{m}}{\lambda}
\end{equation}
fixed. Note that if the limit of $\bnabla_{\bx}G(\bx,\bx')$ exists then necessarily $G=O(1)$ in the outer region, at least up to an immaterial constant, which without loss of generality we take to be independent of $\lambda$. In contrast, in the inner region, which shrinks in size as $\lambda\to0$, the gradient may very well be singular in $\lambda$. 

Consider first the inner region, where $\bb{Y}=O(1)$. Writing $G(\bx,\bx')=g(\bb{Y};\vec{m},\lambda)$, the function $g$ satisfies  
\begin{equation}\label{g eq}
\nabla^2_{\bb{Y}}g=-\lambda^{-1}\delta(\bb{Y}+\vec{m})+\frac{\lambda^2}{\mathcal{V}} \quad \text{for} \quad \vec{m}\bcdot \bb{Y}<0,
\end{equation}
which follows from \eqref{G eq} and \eqref{lambda def}, 
along with the Neumann condition
\begin{equation}\label{g bc}
\vec{m}\bcdot\bnabla_{\bb{Y}}g=0 \quad \text{at} \quad  \vec{m}\bcdot \bb{Y}=0,
\end{equation}
which follows from \eqref{G bc}, as well as matching conditions in the limit $\vec{m}\bcdot\bb{Y}\to-\infty$. Given the scaling of the source in \eqref{g eq}, we pose the inner expansion 
\begin{equation}
g(\bb{Y};\vec{m},\lambda) \sim \lambda^{-1}g_{-1}(\bb{Y};\vec{m}) + o(\lambda^{-1}). 
\end{equation}
At leading order, \eqref{g eq} and \eqref{g bc} respectively give 
\begin{equation}\label{gm1 eq}
\bnabla^2_{\bb{Y}}g_{-1}=-\delta(\bb{Y}+\vec{m}), \quad \bb{Y}\bcdot\vec{m}<0
\end{equation}
and
\begin{equation}\label{gm1 bc}
\vec{m}\bcdot\bnabla_{\bb{Y}}g_{-1}=0 \quad \text{at} \quad \bb{Y}\bcdot\vec{m}=0.
\end{equation}
As for matching, the $O(1)$ scaling of the Green's functions in the outer region implies the far-field condition
\begin{equation}\label{gm1 far}
g_{-1}\to0, \quad \bb{Y}\bcdot\vec{m}\to-\infty.
\end{equation}
The solution of the leading-order inner problem \eqref{gm1 eq}--\eqref{gm1 far} is readily obtained by the method of images as
\begin{equation}
g_{-1}(\bb{Y};\vec{m}) = \frac{1}{4\pi|\bb{Y}+\vec{m}|} + \frac{1}{4\pi|\bb{Y}-\vec{m}|}.
\end{equation}

Consider now the outer region, where $\bx=O(1)$. The formulation of the problem in the outer region is the same as the original problem \eqref{G eq}--\eqref{G bc}, except that the Neumann condition \eqref{G bc} in general excludes the point $\vec{m}$ on the boundary, where matching with the inner region holds instead. Note, in particular, that the Dirac delta forcing in \eqref{G eq} disappears into the latter singular point. Thus, posing the expansion
\begin{equation}
G(\bx,\bx') \sim G_0(\bx;\vec{m}) + o(1), 
\end{equation}
at leading order we have 
\begin{equation}\label{G0 eq}
\nabla^2_{\bx}G_0 = \frac{1}{\mathcal{V}}, \quad \bx\in B(\vec{0}),
\end{equation}
with
\begin{equation}\label{G0 bc}
\bn\bcdot\bnabla_{\bx}G_0 = 0, \quad \bx\in \partial B(\vec{0})\big{/}\vec{m},
\end{equation}
as well as the matching condition 
\begin{equation}\label{G0 match}
G_0(\bx;\vec{m}) \sim \frac{1}{2\pi|\bx-\vec{m}|} \quad \text{as} \quad \bx\to\vec{m}. 
\end{equation}
Note that the gradient of the outer Green's function is precisely the quantity we set out to calculate, i.e.,
\begin{equation}\label{requisite grad}
\underset{\bx'\to\vec{m}}{\lim} \bnabla_{\bx} G(\bx;\bx') = \bnabla_{\bx} G_0(\bx;\vec{m}).
\end{equation}
This quantity is now evaluated by solving the boundary-value problem \eqref{G0 eq}--\eqref{G0 match}, which is clearly independent of the auxiliary parameter $\lambda$ and no longer involves any limiting process or scale separation. 

Practically, it is more convenient to consider the problem governing the even extension of $G_0(\bx;\vec{m})$ about the face centered at $\bx=\vec{m}$. The extended  domain consists of the inclusion of $B(\vec{0})$, $B(\vec{m})$, and the symmetry face; let's denote this rectangular domain $B^2_{\vec{m}}$ and its boundary $\partial B^2_{\vec{m}}$. The evenly extended problem consists of the Poisson equation
\begin{equation}\label{rect eq}
\nabla^2_{\bx}G_0 = \frac{1}{\mathcal{V}}-2\delta(\bx-\vec{m}), \quad \bx\in B^2_{\vec{m}},
\end{equation}
where note that $\mathcal{V}$ still denotes the volume of a single box rather than the volume of the extended domain, 
together with the Neumann condition
\begin{equation}\label{rect bc}
\bn\bcdot\bnabla_{\bx}G_0 = 0, \quad \bx\in \partial B^2_{\vec{m}}.
\end{equation}
Clearly the solution to the above problem is even about the three symmetry planes of the rectangular domain. This, together with the Neumann condition \eqref{rect bc}, implies that the solution can be extended periodically such that the singular points are at the vertices of the rectangular lattice 
\begin{equation}
\bx=\vec{m}+ \bd(\vec{q}), \quad \bd(\vec{q})=4q_1\bbb_1+2q_2\bbb_2 + 2q_3 \bbb_3,
\end{equation}
where $\{q_1,q_2,q_3\}\in\mathbb{Z}^n$ and
\begin{equation}
\bbb_1=\vec{m}, \quad \bbb_2=\be_{\text{mod}(l+1,3)}, \quad \bbb_3=\be_{\text{mod}(l+2,3)},
\end{equation}
the latter three unit vectors forming a right-handed orthogonal basis. This periodicity suggests writing the solution  as the Fourier series
\begin{equation}
G_0(\bx,\vec{m})=\sum_{\br} A_{\br} e^{i 2\pi\br \cdot (\bx-\vec{m})}, \quad \bx\in B^2_{\vec{m}},
\label{eq:ansatz_periodic}
\end{equation}
where the sum is over the set of reciprocal lattice vectors 
\begin{equation}
\br(\vec{q}) =\frac{q_1}{4} \bbb_1+\frac{q_2}{2} \bbb_2 + \frac{q_3}{2} \bbb_3.
\end{equation}
Substituting the ansatz \eqref{eq:ansatz_periodic} in the governing equation \eqref{rect eq} yields
\begin{equation}
A_{\br}=\frac{1}{32\pi^2|\br|^2} \quad \text{for} \quad \br \neq \bzero,
\end{equation}
whereas the coefficient $A_{\bzero}$ can be set to zero on account of the additive freedom of the Neumann Green's function. We thereby find the formal solution
\begin{equation}\label{G series} 
G_0(\bx;\vec{m})=\frac{1}{32\pi^2}\sum'_{\br}{}{} \frac{e^{i 2\pi \br \cdot (\bx-\vec{m})}}{|\br|^2},
\end{equation} 
where the dash on the summation indicates that the term $\br = \bzero $ is to be omitted. The series \eqref{G series} is conditionally convergent. To meaningfully transform it into an absolutely convergent series that can be efficiently computed and differentiated we use results developed in \cite{nijboer57}. Thus, noting that \eqref{G series} is a special case of Eq.~[22] in \cite{nijboer57}, up to a multiplicative constant and differences in notation, we find that 
\begin{multline}
G_0(\bx;\vec{m})=\frac{1}{32\pi^2}\sum'_{\br}{}{}\frac{\Gamma (1,\pi|\br|^2)}{|\br|^2} e^{i 2\pi \br \cdot (\bx-\vec{m}) } - \frac{1}{32\pi}\\+\frac{\mathcal{V}}{16\pi^{3/2}}\sum_{\bd} \frac{\Gamma(1/2,\pi|\bd-\bx+\vec{m}|^2)}{|\bd-\bx+\vec{m}|},
\end{multline} 
where $\Gamma(n,x)$ is the incomplete Gamma function. The requisite gradient \eqref{requisite grad} can now be obtained by term-by-term differentiation:
\begin{align}\label{grad result}
\nabla_{\bx} G_0(\bx;\vec{m})=&\frac{1}{32\pi^2}\sum'_{\br}{}{} \frac{\Gamma (1,\pi|\br|^2)}{|\br|^2} i 2\pi \br e^{i 2\pi \br \cdot (\bx-\vec{m})  } + \\
& \frac{\mathcal{V}}{16\pi}\sum_{\bd} \frac{\bd -\bx+\vec{m}}{|\bd-\bx+\vec{m}|^2}\bigg[2e^{-\pi |\bd-\bx+\vec{m}|^2}+\frac{\Gamma(1/2,\pi|\bd-\bx+\vec{m}|^2)}{\sqrt{\pi}|\bd-\bx+\vec{m}|}\bigg]. \notag
\end{align}
In this form, the first few terms suffice to achieve convergence. We have verified that the results obtained using \eqref{grad result} agree with finite elements simulations \cite{hecht12a} of \eqref{rect eq}--\eqref{rect bc}.

\bibliographystyle{siamplain}
\bibliography{references}

\end{document}